\documentclass[a4paper,twocolumn,english,aps,prl,superscriptaddress]{revtex4-1}
\usepackage[T1]{fontenc}
\usepackage[latin9]{inputenc}
\usepackage{color}
\usepackage{babel}
\usepackage{bm}
\usepackage{amsmath}
\usepackage{amssymb}
\usepackage{graphicx}
\usepackage[unicode=true,pdfusetitle,
 bookmarks=true,bookmarksnumbered=false,bookmarksopen=false,
 breaklinks=false,pdfborder={0 0 1},backref=false,colorlinks=false]
 {hyperref}
\usepackage{breakurl}

\makeatletter

\usepackage{babel}
\usepackage{babel}

\makeatother
\begin{document}
\title{Higher-order Fabry-P\'erot Interferometer from Topological
Hinge States}
\author{Chang-An Li}
\email{changan.li@uni-wuerzburg.de}

\affiliation{Institute for Theoretical Physics and Astrophysics, University of
W\"urzburg, 97074 W\"urzburg, Germany}
\author{Song-Bo Zhang}
\email{song-bo.zhang@uni-wuerzburg.de}

\affiliation{Institute for Theoretical Physics and Astrophysics, University of
W\"urzburg, 97074 W\"urzburg, Germany}
\author{Jian Li}
\affiliation{School of Science, Westlake University, 18 Shilongshan Road, Hangzhou
310024, Zhejiang Province, China}
\affiliation{Institute of Natural Sciences, Westlake Institute for Advanced Study,
18 Shilongshan Road, Hangzhou 310024, Zhejiang Province, China}
\author{Björn Trauzettel}
\affiliation{Institute for Theoretical Physics and Astrophysics, University of
W\"urzburg, 97074 W\"urzburg, Germany}
\date{\today}
\begin{abstract}
We propose an intrinsic 3D Fabry-P\'erot type interferometer,
coined \textquotedblleft higher-order interferometer\textquotedblright ,
that utilizes the chiral hinge states of second-order topological
insulators and \textit{cannot} be equivalently mapped to 2D space
because of higher-order topology. Quantum interference
patterns in the two-terminal conductance of this interferometer are
controllable not only by tuning the strength but also, particularly,
by rotating the direction of the magnetic field applied perpendicularly
to the transport direction. Remarkably, the conductance exhibits
a characteristic beating pattern with multiple frequencies with respect
to field strength or direction. Our novel interferometer provides
feasible and robust magneto-transport signatures to probe the particular
hinge states of higher-order topological insulators.
\end{abstract}
\maketitle
\textit{\textcolor{blue}{Introduction.}}\textit{\textemdash }Higher-order
topological insulators (HOTIs) feature gapless excitations, similar
to traditional (first-order) topological insulators, that are protected
by bulk electronic topology but localized at open boundaries at least
two dimensions lower than the insulating bulk \citep{Benalcazar17Science,BBH17prb,Slager15prb,PengY17prb,Langbehn17prl,SongZD17prl, Schindler18SA,Geier18prb,Ezawa18prl,Khalaf18prb,ParkMJ19prl,Trifunovic19prx,YouYZ18prb,Hirosawa20prl,Franca18prb,Miert18prb}.
For instance, 3D second-order topological insulators (SOTIs) host
1D chiral or helical states along specific hinges of the systems.
In recent years, HOTIs have triggered widespread research interest,
owing to their discoveries in a variety of candidate systems, promotion
of our understanding of topological states of matter, and potential
applications \citep{Schindler18NP,Imhof18np,Peterson18nature,Serra-Garcia18nature,ChenXD19prl,PengY19prl,Ghosh20prb, Hassan19NPho,Ni19nm,XieBY19prl,QiY20prl,Roy19prb,Andras20prb,XLSheng19prl,LiC20prb,LiCA20prl, LiHQ20prl,XYZhu18prb,XWLuo19prl,Ezawa19prb,SBZhang20prb-braiding,SBZhang20PRR,Pahomi20prr}.
So far, most efforts have been put into the potential realization
and electronic characterization of HOTIs. However, the transport properties
of HOTIs remain largely unexplored, despite of a few works associated
with superconductivity \citep{Queiroz19prl,LiCZ20prl,Choi20nm}. Indeed,
for 3D SOTIs, an intriguing open question is whether the emergent
hinge states can exhibit any particular phenomena in normal-state
transport.

\begin{figure}
\includegraphics[width=1\columnwidth]{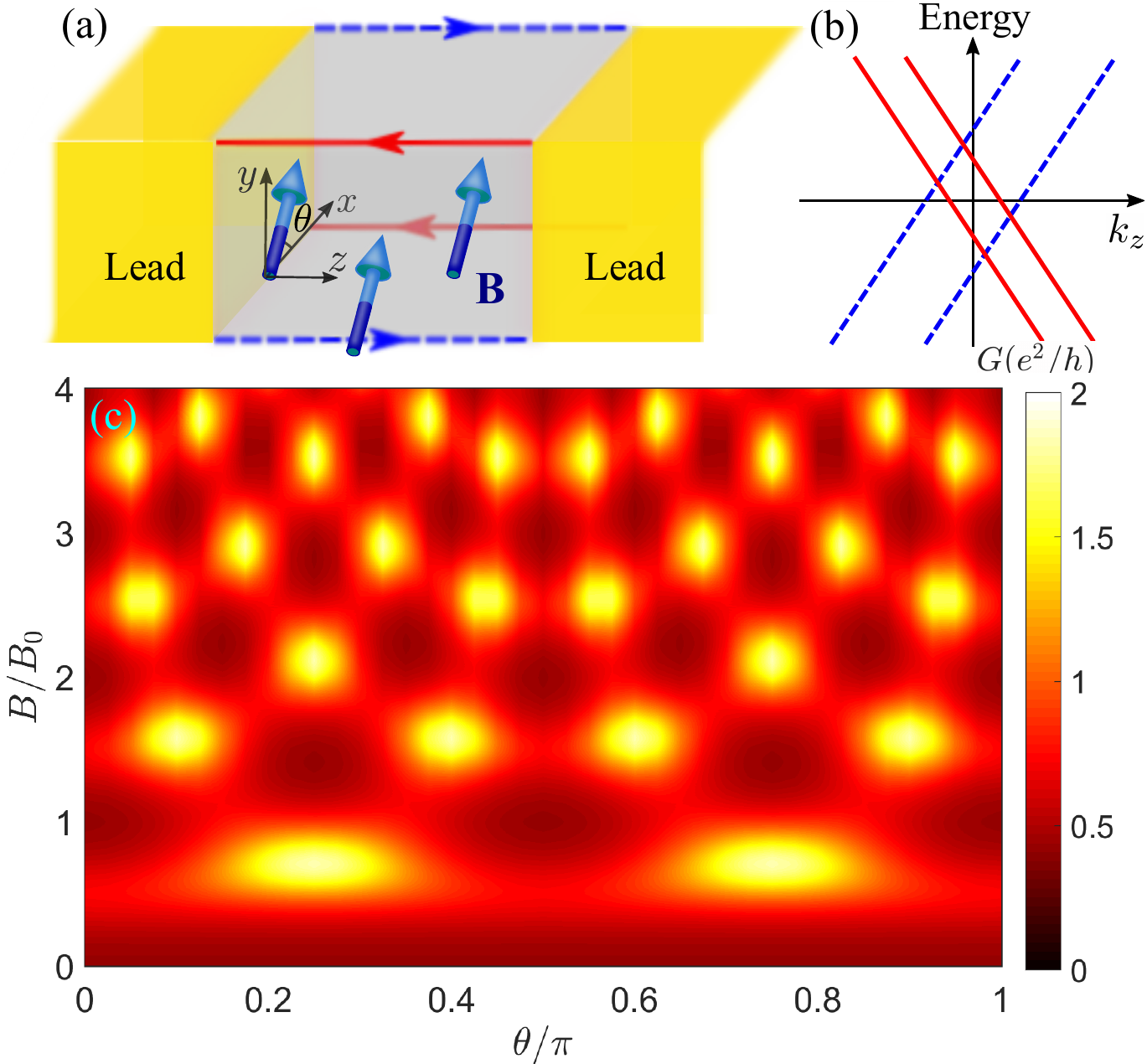}

\caption{(a) Schematic of the higher-order interferometer: a SOTI with four
chiral hinge states (solid red and dashed blue lines) are connected
to two leads (yellow). Adjacent chiral hinge states form interference
loops in the presence of finite reflections at the interfaces. A magnetic
field ${\bf B}$ perpendicular to $z$-direction is applied to the
SOTI (gray). (b) The hinge states have linear dispersion and are shifted
in $k_{z}$-direction by ${\bf B}$. (c) Density plot of conductance
with respect to the field strength $B$ and direction $\theta$. $B_{0}=\phi_{0}/S_{f}$
with $\phi_{0}$ the flux quantum and $S_{f}$ the area of the front
surface of the SOTI.}

\label{Fig: setup}
\end{figure}

One appealing route towards this question involves interferometers
built of SOTIs, which enable us to study quantum-coherent transport
of hinge states. Propagating hinge states that form interference loops
enclosing a magnetic flux applied to the system pick up an Aharonov-Bohm
(AB) phase \citep{Aharonov59PR}. In presence of quantum coherence,
the AB phase will give rise to quantum oscillations in transport characteristics
such as the charge conductance. Quantum interference patterns in the
two-terminal conductance have been employed to detect topological
phases of matter, for instance, surface states of topological insulators
\citep{Bardarson10prl,ZhangY10prl,PengH10nm,Bardarson13rpp}, chiral
Majorana modes \citep{Akhmerov09prl,Fu09prl,LiJ12prb,LiCA19prb},
and topological Dirac semimetals \citep{LXWang16NC}.

In this work, we propose a higher-order Fabry-P\'erot interferometer
to probe hinge states of SOTIs. Our basic setup, shown in Fig.\ \ref{Fig: setup}(a),
is composed of a rectangular chiral SOTI in contact with two leads. The chiral hinge states, existing in 3D space, form interference
loops due to finite reflections (not shown) at the two interfaces,
and their energy dispersions split in a non-uniform manner under magnetic
fields, as shown in Fig.\ \ref{Fig: setup}(b). Particular quantum
interference patterns in the two-terminal conductance, arising from
the AB effect as exemplified by Fig.\ \ref{Fig: setup}(c), can be
observed either by tuning the field strength $B$ or direction $\theta$.
In addition, owing to the intrinsic 3D nature of the interferometer,
there are generally two frequencies in the magneto-conductance oscillations,
leading to a beating pattern. These features do not depend on the
details of the junction, such as the electronic spectrum of the leads,
and are stable against disorder and dephasing. Hence, they provide
robust transport signatures of hinge states in 3D SOTIs.

\textit{\textcolor{blue}{General analysis based on scattering matrix
theory.}}\textit{\textemdash }Our proposed interferometer involves
a 3D SOTI with four chiral hinge states attached to two leads in $z$-direction,
as sketched in Fig.\ \ref{Fig: setup}(a). Adjacent chiral hinge
states form interference loops because of finite reflections at the
interfaces, as will be discussed below. A magnetic field ${\bf B}=B(\cos\theta,\sin\theta,0)$
in $x$-$y$ plane is applied in the SOTI region, where $B$ measures
the field strength and $\theta$ the field direction with respect
to $x$-direction.

Before presenting concrete results based on specific models, it is
instructive to analyze the main transport features of the interferometer
using a phenomenological scattering matrix approach \citep{Buttiker92prb,Nazarovbook,Maciejko10prb}.
The transport properties of the setup are encoded in a scattering
matrix that directly connects the conducting channels in the left
and right leads. The scattering processes at the
two interfaces between the leads and the SOTI can be described by
two scattering matrices, respectively. Each matrix consists of four
components: transmission from left to right $t_{L/R}$, transmission
from right to left $t_{L/R}'$, reflection from the right $r_{L/R}$
and reflection from the left $r_{L/R}'$, where the subscript ($L$
and $R$) distinguishes the left and right surfaces. At low energies,
the only conducting channels in the SOTI are the four hinge states
which have linear dispersion and are localized at the four different
hinges of the cuboid. In the presence of a magnetic field, their
propagation in the SOTI will pick up AB phases that can be described
by a phase matrix $U\equiv e^{2i\lambda}e^{i\varphi\sigma_{z}\otimes\sigma_{0}/2}e^{i\phi\sigma_{z}\otimes\sigma_{z}/2}$,
where $\sigma_{z}$ is a Pauli matrix, $\sigma_{0}$ the $2\times2$
identity matrix,
\begin{equation}
\varphi=BLW_{x}\cos\theta\ \text{and }\phi=BLW_{y}\sin\theta\label{eq:two-phases}
\end{equation}
are the magnetic fluxes threading the two surfaces, respectively,
with $L$ the distance between the two leads and $W_{x/y}$ the widths
of the sample in $x/y$-directions. Moreover, $\lambda=k_{F}L$ is
the dynamical phase with $k_{F}$ the Fermi wave number in the absence
of magnetic fields. By eliminating the scattering amplitudes in the
SOTI region, we derive analytically an effective $2\times2$ scattering
matrix that directly connects the two interfaces \citep{Li2020SM}
\begin{equation}
\mathcal{S}(B,\theta)=\Phi_{+}(e^{-i\lambda}-e^{i\lambda}r_{L'}\Phi_{-}r_{R}\Phi_{+})^{-1},\label{eq:scattering matrix}
\end{equation}
where the phase matrices $\Phi_{\pm}\equiv e^{i(\varphi\pm\phi)\sigma_{z}/2}$
account for the AB phase differences between the two right-moving
and between the two left-moving hinge channels, respectively. At zero
temperature, the two-terminal conductance of the setup can be evaluated
as
\begin{equation}
G(B,\theta)=\frac{e^{2}}{h}\mathrm{Tr}[t_{R}^{\dagger}t_{R}\mathcal{S}(B,\theta)t_{L}t_{L}^{\dagger}\mathcal{S}^{\dagger}(B,\theta)],\label{eq:conductance}
\end{equation}
where $h$ is the Planck constant and $e$ is electron
charge. According to Eqs.\ (\ref{eq:scattering matrix})
and (\ref{eq:conductance}), if there is no transmission across any
of the two interfaces, i.e., $t_{L}=0$ or $t_{R}=0$, then $G$ vanishes.
In the opposite limit, where the interfaces are completely transparent
for the hinge states, $r_{L'}=0$ and $r_{R}=0$, we find that the
matrix $\mathcal{S}$ as well as $t_{R}^{\dagger}t_{R}$ and $t_{L}t_{L}^{\dagger}$
become diagonal. As a result, $G$ becomes quantized at $2e^{2}/h$
and is independent of the magnetic field. These results indicate the
necessary condition for a successful interferometer: non-trivial transmission
and reflection at the two interfaces for the hinge states.

When the interfaces are partially transparent, Eq.\ (\ref{eq:scattering matrix})
indicates the formation of Fabry-P\'erot interference loops. Moreover,
the matrix $\mathcal{S}$ contains explicitly two phases $\varphi\pm\phi$
in general. This indicates the appearance of beating patterns with
two frequencies in the magneto-conductance. Notably, the two frequencies
are intimately connected to the magnetic fluxes threading the different
surfaces of the SOTI. They are solely determined by the geometry of
the sample and insensitive to the details of the interface barriers.
The oscillation pattern of $G$ remains qualitatively the same even
in the presence of a dynamic phase. We verified
these results by properly parametrizing the scattering matrices \citep{Li2020SM}.

\textit{\textcolor{blue}{Model simulation and method.}}\textit{\textemdash }To
demonstrate these features of the interferometer explicitly, we consider
an effective model for chiral SOTIs \citep{Schindler18SA}
\begin{align}
H({\bf k})= & \Big(m+b\sum_{i=x,y,z}\cos k_{i}\Big)\tau_{3}+v\sum_{i=x,y,z}\sin k_{i}\sigma_{i}\tau_{1}\nonumber \\
 & +\Delta(\cos k_{x}-\cos k_{y})\tau_{2},\label{eq:Hamiltonian}
\end{align}
where ${\bf k}=(k_{x},k_{y},k_{z})$ is the wave vector. $\bm{\tau}=(\tau_{1},\tau_{2},\tau_{3})$
and $\bm{\sigma}=(\sigma_{x},\sigma_{y},\sigma_{z})$ are Pauli matrices
acting on orbital and spin spaces, respectively; $m$, $b$, $v$
and $\Delta$ are model parameters. Without loss of generality, we
set the lattice constant and the velocity $v$ to unity hereafter.
When $1<|m/b|<3$ and $\Delta=0$, the model describes 3D topological
insulators with gapless surface states \citep{HJZhang09nphys}. The
surface states are protected by time-reversal symmetry $\mathcal{T}=i\sigma_{2}\mathcal{K}$,
where $\mathcal{K}$ represents complex conjugation. A finite $\Delta\neq0$
breaks time-reversal and $\mathcal{C}_{4}$ rotation (with the rotation
axis pointing in $z$-direction) symmetries individually. It opens
gaps in the surface states. However, the $\Delta$ term preserves
the combined symmetry $\mathcal{C}_{4}\mathcal{T}$, as indicated
by $(\mathcal{C}_{4}\mathcal{T})H(k_{x},k_{y},k_{z})(\mathcal{C}_{4}\mathcal{T})^{-1}=H(k_{y},-k_{x},-k_{z})$.
As a result, the gaps opened by $\Delta$ depend on the surface orientation,
leading to gapless chiral hinge states localized at the hinges connecting
different surfaces.

We take into account the orbital effect of the magnetic field via
the Peierls replacement in the hopping interaction $T_{ij}\rightarrow T_{ij}\exp(2\pi i\int_{r_{i}}^{r_{j}}d{\bf {\bf r}}\cdot{\bf A}/\phi_{0})$,
where $T_{ij}$ is the hopping amplitude from sites $r_{i}$ to $r_{j}$,
$\phi_{0}=h/e$ is flux quantum. ${\bf A}$ is the vector potential
for the magnetic field and it is chosen as ${\bf A}=B(0,0,y\cos\theta-x\sin\theta)$
for concreteness \citep{Li2021footnote1}.

For simplicity, we model the metallic leads with a conventional quadratic
energy dispersion and assume only a few transport channels in both
leads such that considerable reflections for the hinge channels are
generated at the interfaces. Furthermore, we consider the size of
the system to be much larger than the decay length of the hinge states
in order to have a well-defined multiple-loop interferometer based
on hinge states. Under these considerations, we calculate the two-terminal
conductance numerically, employing the standard Landauer-B\"uttiker
approach \citep{Landauer70pm,buttiker_four-terminal_1986,Dattabook}
in combination with lattice Green functions (see the Supplemental
Material \citep{Li2020SM}). We emphasize that our main results illustrated
below remain qualitatively the same if we choose other models for
SOTIs or leads.

\begin{figure}
\centering

\includegraphics[width=1\columnwidth]{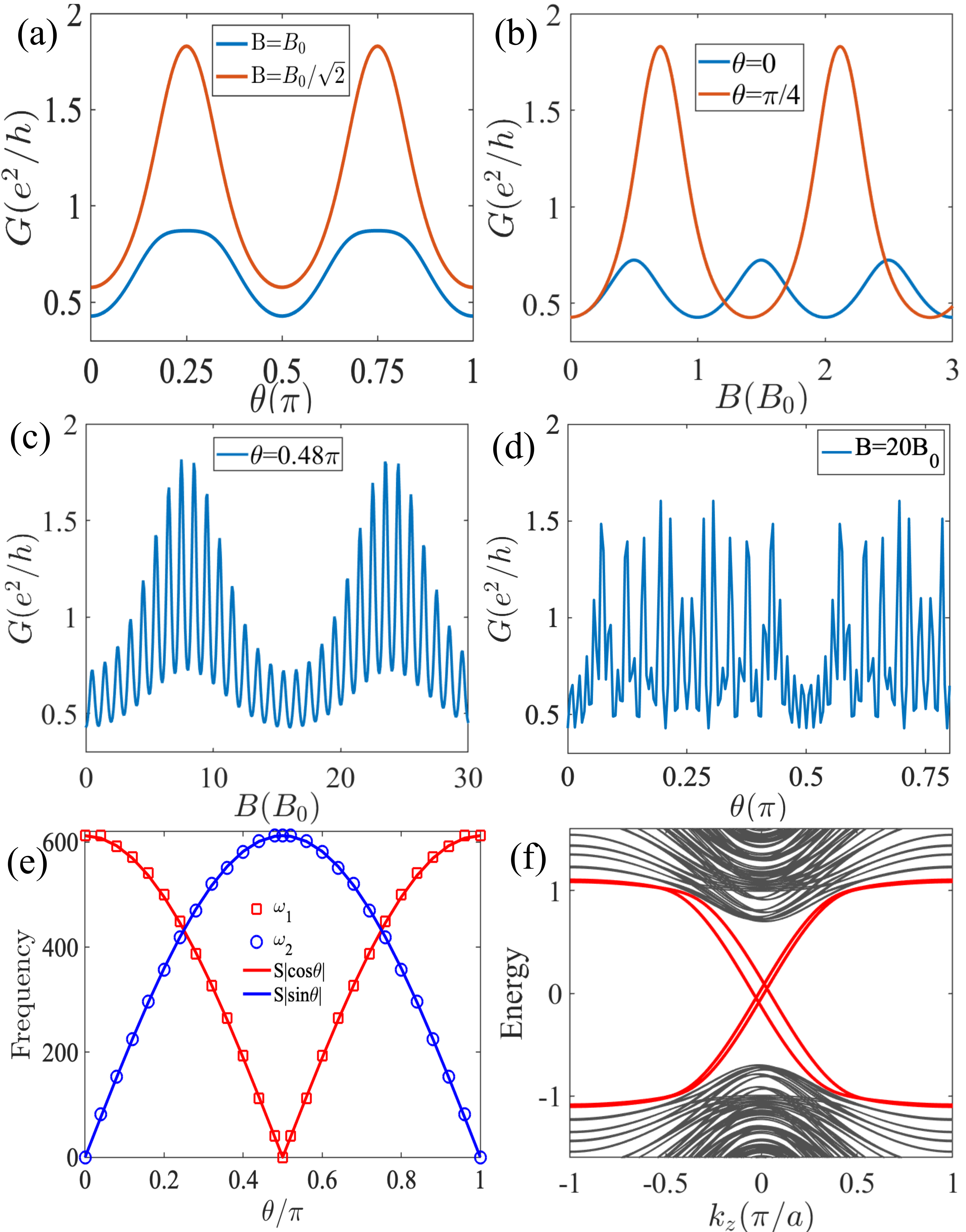}

\caption{(a) Conductance $G$ as a function of field direction $\theta$ at
small field strengths $B=B_{0}$ and $B_{0}/\sqrt{2}$, respectively.
(b) $G$ as a function of $B$ for $\theta=0$ and $\pi/4$, respectively.
In these cases, the oscillations have a single frequency. (c) Particular
beating patterns as varying $B$ at angle $\theta=0.48\pi$. (d) ``Irregular''
beating patterns as varying $\theta$ at a large field strength $B=20B_{0}$.
(e) The extracted frequencies (square and circle dots) as a function
of $\theta$. The two frequencies can be described by $\omega_{1}=S|\cos\theta|$
and $\omega_{2}=S|\sin\theta|$. (f) The low-energy spectrum of the
SOTI in the presence of a magnetic field $B=2B_{0}$ and $\theta=0.15\pi$.
Other parameters are $L_{z}=60a$, $W_{x}=W_{y}=12a$, $m=2,b=-1,v=1$,
$\Delta=1$, and the Fermi energy $E_{F}=0.002$.}

\label{fig:oscillations}
\end{figure}

\textit{\textcolor{blue}{Quantum interference pattern.}}\textit{\textemdash }Now,
we analyze the dependence of the conductance $G$ on the magnetic
field, combining general scattering theory and concrete numerical
simulations. Equation\ (\ref{eq:conductance}) implies an oscillation
pattern of $G$ with respect to the field direction $\theta$. As
shown in Fig.\ \ref{fig:oscillations}(a), $G(\theta)$ is periodic
in $\theta,$ in accordance with the scattering theory. Explicitly,
we find that for weak magnetic fields $B\leq B_{0}$, $G(\theta)$
is approximately a sinusoidal function of $\theta$ and takes the
maximal value at $\theta=\pi/4+n\pi/2$, $n\in\{0,1,2,3\}$, when
$W_{x}=W_{y}$. Here, $B_{0}$ corresponds to the field strength at
which the flux enclosed by the front surface $S_{f}$ is one flux
quantum for $\theta=0$. Thus, $G(\theta)$ has a period of $\pi/2$
in $\theta$. Moreover, $G(\theta)$ is minimal at $\theta=\theta_{c}$
and symmetric in $\theta-\theta_{c}$, where $\theta_{c}=n\pi/2$.
For strong magnetic fields $B>B_{0}$, the number of conductance peaks
increases with increasing $B$, see Fig.\ \ref{Fig: setup}(c). When
$W_{x}\neq W_{y},$ the period in $\theta$ becomes $\pi$ but $G(\theta)$
is still symmetric in $\theta-\theta_{c}$.

Equation\ (\ref{eq:conductance}) also indicates an oscillation pattern
of $G$ with respect to the field strength $B,$ which is again fully
confirmed by our numerical simulations. When the magnetic field is
applied in $x$- or $y$-directions, or at the specific angle $\theta=\pm\text{arctan}(W_{x}/W_{y})$,
$G(B)$ exhibits simple oscillations with a single
frequency, see Fig.\ \ref{fig:oscillations}(b). Generally, the oscillating
conductance takes maximal or minimal values when the interference
loop encloses half a flux quantum. In our cases, $G(B)$ takes maximal
values at odd multiples of $B_{0}/2$ for $\theta=0$. The oscillation
amplitude is relatively smaller since only two of the four loops enclose
half a flux quantum at this field direction. For $\theta=\pi/4$,
$G(B)$ takes maximal values at odd multiples of $B_{0}/\sqrt{2}$,
where the interference loop also encloses half a flux quantum, leading
to a resonance peak of $G(B)$. These features signify the interferometer
formed by hinge states being of Fabry-P\'erot type, as we further
explain below.

Notably, there exist beating patterns, as signified by Eq.\ (\ref{eq:scattering matrix}),
where the matrix $\mathcal{S}$ explicitly contains the two phases
$\varphi\pm\phi$. When the magnetic field deviates away from the
special directions at $\theta=n\pi/2$ (with $n\in\{0,1,2,3\}$) and
$\pm\text{arctan}(W_{x}/W_{y})$, beating oscillations of $G(B)$
are clearly observed, as shown in Fig.\ \ref{fig:oscillations}(c).
By performing discrete Fourier transformation to the beating patterns,
we obtain precisely two frequencies $\omega_{1}$ and $\omega_{2}$.
These frequencies depend strongly on the field direction $\theta$
{[}dotted lines in Fig.\ \ref{fig:oscillations}(e){]}. Explicitly,
we find that the two frequencies can be well described by $\omega_{1}=|S\cos\theta|=|\phi|/B$
and $\omega_{2}=|S\sin\theta|=|\varphi|/B$ {[}solid lines in Fig.\ \ref{fig:oscillations}(e){]},
respectively, where $S$ is the area of the surfaces of the system
(we consider the case with $W_{x}=W_{y}$ for simplicity). This corresponds
exactly to the two AB phases in Eq.\ (\ref{eq:two-phases}), in excellent
agreement with the results obtained from scattering-matrix analysis.
When $\theta=n\pi/2$, only one of the two frequencies survives. When
$\theta=\pi/4+n\pi/2$, the two frequencies become identical. In both
cases, the beating behavior in the oscillations disappear. Similarly,
$G(\theta)$ also shows beating-like patterns with respect to the
field direction $\theta$ with irregular peaks and dips for large
magnetic fields $B\gg B_{0}$, as shown in Fig.\ \ref{fig:oscillations}(d).
This direction-induced beating behavior is another manifestation of
the two AB phases.

\textit{\textcolor{blue}{Higher-order Fabry-P\'erot interference.}}\textit{\textemdash }Next,
we clarify, in which sense our quantum interference pattern is a higher-order
Fabry-P\'erot type interference. The two frequencies in the beating
patterns correspond physically to two areas of interference loops.
As rotating the magnetic field, the two frequencies match the effective
areas of front surface $|S\cos\theta|$ and top surface $|S\sin\theta|$
quite well, see Fig.\ \ref{fig:oscillations}(e). This fact indicates:
(i) the adjacent hinge states with opposite chirality form effective
interference loops and the interference is typically of Fabry-P\'erot
type; and (ii) there are totally four interference loops but any two
opposite surfaces of the sample (namely, the front and back surfaces,
or the top and bottom surfaces) have the same effective area because
of the chosen symmetry of the system \citep{Li2021footnote2}. The
interference loops are made of chiral hinge modes located in 3D space,
protected by higher-order topology. When rotating
the magnetic field, one of the frequencies increases, whereas the
other one decreases. Moreover, the ratio between the two frequencies depends
on $\theta$ as $S_{f}/S_{t}=W_{y}|\cot\theta|/W_{x}$. Thus, the
two frequencies coincide at the critical field directions $\theta_{c}=\text{arctan}(W_{y}/W_{x})$
and $\pi-\text{arctan}(W_{y}/W_{x})$, as shown in Fig.\ \ref{fig:oscillations}(e).
These features indicate the 3D nature of the interferometer.

The mechanism of the interferometer can be better understood by analyzing
the splitting of hinge states under magnetic fields. In the absence
of magnetic fields, the four chiral hinge states have a double degenerate
linear spectrum in $k_{z}$-direction, i.e., $\pm vk_{z}$. The magnetic
field gives rise to a spatially varying vector potential. Note that
the hinge states are localized at different hinges of the system.
The local vector potential splits the linear spectrum of the hinge
states. Under the chosen gauge, the spectra of hinge states are split
as $+v(k_{z}\pm\delta k_{1}^{z})$ and $-v(k_{z}\pm\delta k_{2}^{z})$,
where the splittings are determined by $\delta k_{1}^{z}=BW_{x}|\sin(\theta-\pi/4)|/2$
and $\delta k_{2}^{z}=BW_{y}|\cos(\theta+\pi/4)|$/2 \citep{Li2020SM}.
Thus, the hinge states acquire finite momenta even for vanishing Fermi
energy {[}Fig.\ \ref{fig:oscillations}(f){]}. When propagating across
the SOTI region, the hinge channels pick up extra phases, $\pm\delta k_{1/2}^{z}L_{z}$.
Such phases turn out to be exactly the AB phases $\phi$ and $\varphi$,
stemming from the magnetic flux enclosed by each loop. Explicitly,
the flux enclosed by front surface $S_{f}$ and top surface $S_{t}$
of the central SOTI region in Fig. \ref{Fig: setup}(a) are given
by $(\delta k_{1}^{z}+\delta k_{2}^{z})L_{z}$ and $2\delta k_{1}^{z}L_{z}$,
respectively. Plugging $\phi,\varphi=\delta k^{z}L_{z}$ into Eq.\ (\ref{eq:scattering matrix}),
this indicates that the dependence of $G$ on ${\bf B}$ can be attributed
to the higher-order Fabry-P\'erot interference of the four hinge
states. At special values of $\theta$, say $\theta=\pi/4$ or $5\pi/4$,
one kind of splitting vanishes whereas the other one remains, $\delta k_{1}^{z}=0$
and $\delta k_{2}^{z}\neq0$ (similar results occur for $\theta=+3\pi/4,-\pi/4$).
In these cases, we have only one frequency.

\begin{figure}
\centering

\includegraphics[width=1\columnwidth]{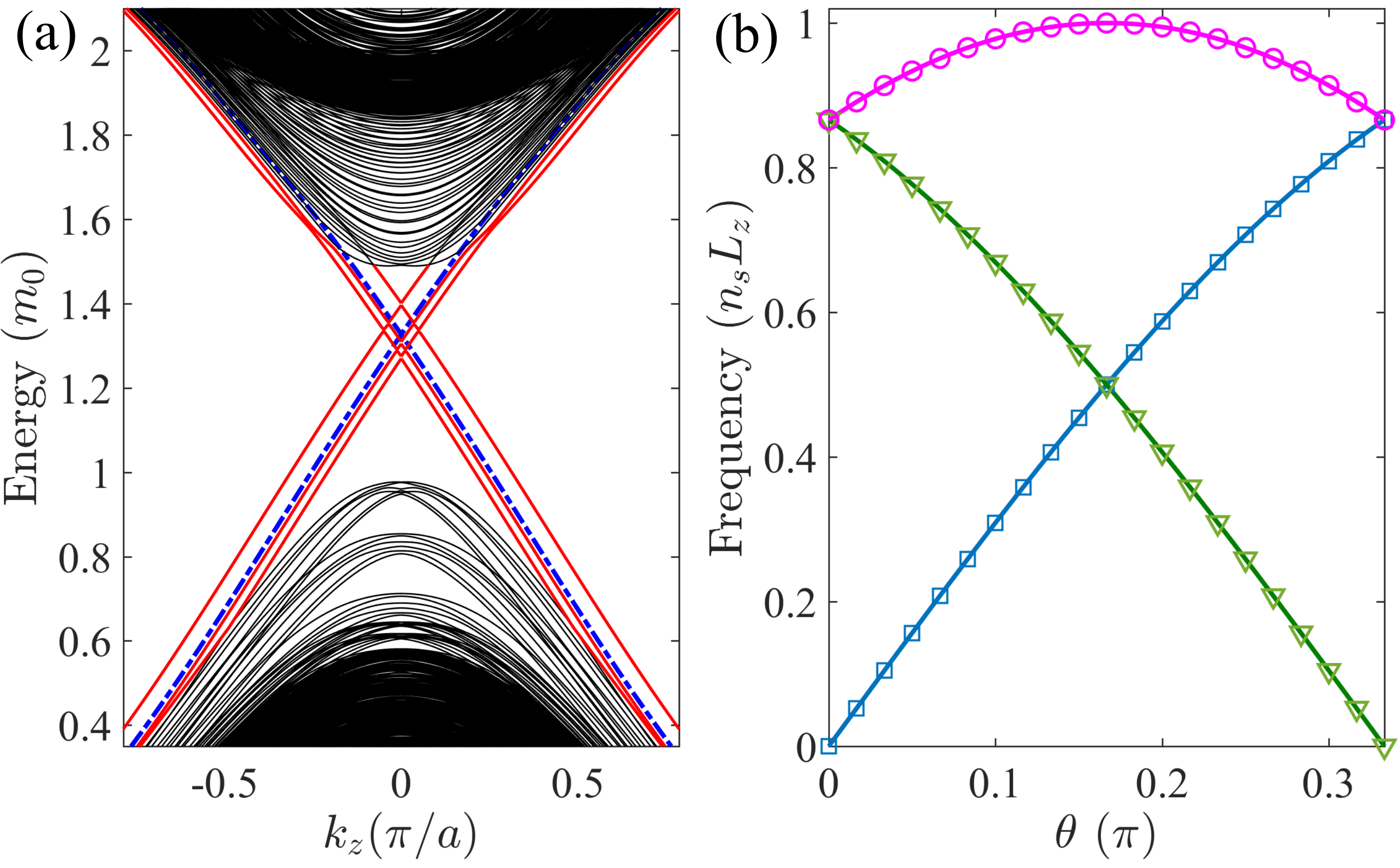}

\caption{(a) Low-energy spectrum of the SOTI in the presence of a magnetic
field. (b) The three frequencies in the case of $\mathcal{C}_{6}$
symmetric SOTIs as functions of $\theta$. The lattice model and related
parameters can be found in the Supplemental Material \citep{Li2020SM}.}

\label{fig:oscillations-Hexagon}
\end{figure}

\textit{\textcolor{blue}{Generalization to $\mathcal{C}_{6}$
symmetric SOTIs.}}\textit{\textemdash }So far, we have focused on
the case of chiral SOTIs with four hinge states and a sample with
(effective) $\mathcal{C}_{4}$ symmetry. However, our scattering theory
can be generalized and applied to SOTIs with more pairs of hinge states.
As an example, we consider a $\mathcal{C}_{6}$ symmetric SOTI with
three pairs of chiral hinge states \citep{ZhangRX20prl} and show
the spectrum in Fig.\ \ref{fig:oscillations-Hexagon}(a). The geometry
considered here is a hexagonal prism with $\mathcal{C}_{6}$ symmetry
in $x$-$y$ plane. The hinge states split generally with different
amounts of momenta under a magnetic field. If we consider an interferometer
similar to the setup in Fig.\ \ref{Fig: setup}(a), we can also observe
characteristic oscillations and beating patterns in the conductance
which depend sensitively on the field direction $\theta$. In this
case, the conductance $G(\theta)$ is $\pi/3$ periodic in $\theta$.
Since there are three pairs of counter-propagating hinge states, the
oscillations can exhibit three frequencies in general \citep{Li2020SM}.
Figure\ \ref{fig:oscillations-Hexagon}(b) illustrates the three
frequencies as a function of field direction $\theta$. Particularly,
the oscillations are described by a single and two frequencies for
$\theta=0$ and $\theta=\pi/6$, respectively.

\textit{\textcolor{blue}{Discussion and summary.}}\textit{\textemdash }In
realistic samples, disorder and dephasing \citep{JiangH09prl} due
to environmental noises may be detrimental to the interference pattern
of hinge states. However, we show numerically that the oscillation
patterns of the conductance in our setups persist under weak disorder
and dephasing \citep{Li2020SM}. This indicates the robustness of
our proposal. Our results based on chiral SOTIs can also be applied
to helical SOTIs, which can be regarded as two copies of chiral SOTIs
related by time-reversal symmetry. Recently, SOTIs have been proposed
in many candidate materials. Among these candidates, bismuth \citep{Schindler18NP}
and axion insulators including EuIn$_{2}$As$_{2}$ and MnBi$_{2}$Te$_{4}$
\citep{XuY19prl,chenR20arxiv} provide potential platforms to test
our predictions.

In summary, we have proposed a higher-order Fabry-P\'erot interferometer
and revealed unique Aharonov-Bohm oscillations arising from topological
hinge states by tuning either strength or direction of an applied
magnetic field. Due to higher-order topology, the interferometer is
intrinsically three-dimensional and features particular beating patterns
in the magneto-conductance. Our results are robust and provide unique
transport signatures of hinge states in higher-order topological insulators.

\begin{acknowledgements}This work was supported by the DFG (SPP1666
and SFB1170 ``ToCoTronics''), the W\"urzburg-Dresden Cluster of
Excellence ct.qmat, EXC2147, project-id 390858490, and the Elitenetzwerk
Bayern Graduate School on ``Topological Insulators''. JL acknowledges support by NSFC under Grants No. 11774317. \end{acknowledgements}

%\bibliographystyle{apsrev4-1-etal-title}
% \bibliography{Refsdata}

\begin{thebibliography}{66}
\bibitem [{Li2({\natexlab{b}})}]{Li2021footnote1}%
  \BibitemOpen
  \href@noop {} {\bibfield  {journal} {\bibinfo  {journal} {We ignore the
  Zeeman interaction in this model because it is expected to have a minor
  influence on the AB effect}\ }}\BibitemShut {NoStop}%
\end{thebibliography}


\appendix
\numberwithin{equation}{section}\setcounter{figure}{0}\global\long\def\thefigure{S\arabic{figure}}
\global\long\def\thesection{S\arabic{section}}
\global\long\def\thesubsection{\Alph{subsection}}

\begin{widetext}
\begin{center}
\textbf{\large{}Supplemental material for ``Higher-order Fabry-P\'erot Interferometer from Topological
Hinge States''}{\large{} }
\par\end{center}{\large \par}
%\tableofcontents

\section{General scattering matrix analysis}

In this section, we present the details for the scattering matrix
analysis of the setup in the main text. Suppose there are $p_{L/R}$
conducting modes at the Fermi level in the left/right leads. We can
define generally $(p_{L/R}+2)\times(p_{L/R}+2)$ scattering matrices,
$S_{L}$ and $S_{R}$, to describe the scattering at the left and
right interfaces, respectively,
\begin{align}
\begin{pmatrix}b_{L}\\
b_{L'}
\end{pmatrix} & =S_{L}\begin{pmatrix}a_{L}\\
a_{L'}
\end{pmatrix},\ \ S_{L}=\begin{pmatrix}r_{L} & t_{L'}\\
t_{L} & r_{L'}
\end{pmatrix},\label{eq:L-matrix}\\
\begin{pmatrix}b_{R'}\\
b_{R}
\end{pmatrix} & =S_{R}\begin{pmatrix}a_{R'}\\
a_{R}
\end{pmatrix},\ \ S_{R}=\begin{pmatrix}r_{R} & t_{R'}\\
t_{R} & r_{R'}
\end{pmatrix}.\label{eq:R-matrix}
\end{align}
Here, $a_{L/R}$ and $b_{L/R}$ indicate the incoming and outgoing
modes that propagate in the leads and scatter at the left/right interface,
respectively; $a_{L'/R'}$ and $b_{L'/R'}$ indicate the incoming
and outgoing hinge modes that propagate in the SOTI and scatter at
the left/right interface, respectively. The scattering matrix $S_{L/R}$
consists of four components $t_{L/R},$ $t'_{L/R},$ $r_{L/R},$ and
$r'_{L/R}$, corresponding to the transmission from left to right,
transmission from right to left, reflection from the right, and reflection
from the left, respectively. In the center SOTI region, the conducting
chiral hinge states pick up an AB phases when applying an external
magnetic field. Thus, the incoming and outgoing modes in the SOTI
can be connected by a phase matrix as

\begin{align}
\begin{pmatrix}a_{R'1}\\
a_{R'2}\\
a_{L'1}\\
a_{L'2}
\end{pmatrix} & =e^{i\lambda/2}\begin{pmatrix}e^{i(\varphi+\phi)/2} & 0 & 0 & 0\\
0 & e^{-i(\varphi+\phi)/2} & 0 & 0\\
0 & 0 & e^{i(\varphi-\phi)/2} & 0\\
0 & 0 & 0 & e^{-i(\varphi-\phi)/2}
\end{pmatrix}\begin{pmatrix}b_{L'1}\\
b_{L'2}\\
b_{R'1}\\
b_{R'2}
\end{pmatrix},\label{eq:HOTI-AB-phase}
\end{align}
where $\lambda=k_{F}L$ is the dynamic phase with $k_{F}$ the Fermi
wave number in $k_{z}$-direction and $L$ the length of the SOTI,
and the two phases are given by
\begin{equation}
\phi=BS_{1}\sin\theta,\ \ \varphi=BS_{2}\cos\theta,
\end{equation}
with $\theta$ the angle between the magnetic field direction and
$x$-axis.

Substituting Eq.\ \eqref{eq:HOTI-AB-phase} into Eq.\ \eqref{eq:R-matrix},
we obtain
\begin{align}
\begin{pmatrix}e^{-i\lambda/2}\Phi_{-}^{\dagger}a_{L'}\\
b_{R}
\end{pmatrix} & =S_{R}\begin{pmatrix}e^{i\lambda/2}\Phi_{+}b_{L'}\\
a_{R}
\end{pmatrix},\label{eq:R-matrix-HOTI}
\end{align}
where $\Phi_{\pm}\equiv e^{i(\varphi\pm\phi)\sigma_{z}/2}$ and the
Pauli matrix $\sigma_{z}$ acts on (pseudo-)spin space for two left-
or right-moving hinge states. Writing Eqs.\ \eqref{eq:R-matrix-HOTI}
explicitly, we have
\begin{align}
e^{-i\lambda/2}\Phi_{-}^{\dagger}a_{L'} & =e^{i\lambda/2}r_{R}\Phi_{+}b_{L'}+t_{R'}a_{R},\label{eq:1-3-HOTI}\\
b_{R} & =e^{i\lambda/2}t_{R}\Phi_{+}b_{L'}+r_{R'}a_{R}.\label{eq:1-4-HOTI}
\end{align}
From Eq.\ \eqref{eq:1-3-HOTI}, we find
\begin{equation}
a_{L'}=e^{i\lambda}\Phi_{-}r_{R}\Phi_{+}b_{L'}+e^{i\lambda/2}\Phi_{-}t_{R'}a_{R}.\label{eq:1-5-HOTI}
\end{equation}
Writing Eq.\ \eqref{eq:L-matrix} explicitly, we have
\begin{align}
b_{L} & =r_{L}a_{L}+t_{L'}a_{L'},\label{eq:1-1}\\
b_{L'} & =t_{L}a_{L}+r_{L'}a_{L'},\label{eq:1-2}
\end{align}
Plugging Eq.\ \eqref{eq:1-5-HOTI} into Eq.\ \eqref{eq:1-2}, we
obtain
\begin{equation}
b_{L'}=t_{L}a_{L}+r_{L'}(e^{i\lambda}\Phi_{-}r_{R}\Phi_{+}b_{L'}+e^{i\lambda/2}\Phi_{-}t_{R'}a_{R}),
\end{equation}
and hence,
\begin{equation}
b_{L'}=(1-e^{i\lambda}r_{L'}\Phi_{-}r_{R}\Phi_{+})^{-1}(t_{L}a_{L}+e^{i\lambda/2}r_{L'}\Phi_{-}t_{R'}a_{R}).\label{eq:Bl-solution-1}
\end{equation}
Plugging this result into Eq.\ \eqref{eq:1-4-HOTI}, we find $b_{R}$
as
\begin{align}
b_{R} & =t_{R}e^{i\lambda/2}\Phi_{+}(1-e^{i\lambda}r_{L'}\Phi_{-}r_{R}\Phi_{+})^{-1}(t_{L}a_{L}+e^{i\lambda/2}r_{L'}\Phi_{-}t_{R'}a_{R})+r_{R'}a_{R}\nonumber \\
 & =e^{i\lambda/2}t_{R}\Phi_{+}(1-e^{i\lambda}r_{L'}\Phi_{-}r_{R}\Phi_{+})^{-1}t_{L}a_{L}+[e^{i\lambda}t_{R}\Phi_{+}(1-e^{i\lambda}r_{L'}\Phi_{-}r_{R}\Phi_{+})^{-1}r_{L'}\Phi_{-}t_{R'}+r_{R'}]a_{R}.\label{eq:result1}
\end{align}
Plugging Eq.\ \eqref{eq:1-2} into Eq.\ \eqref{eq:1-3-HOTI}, we
obtain
\begin{equation}
e^{-i\lambda/2}\Phi_{-}^{\dagger}a_{L'}=e^{i\lambda/2}r_{R}\Phi_{+}(t_{L}a_{L}+r_{L'}a_{L'})+t_{R'}a_{R},
\end{equation}
and hence,
\begin{equation}
a_{L'}=(1-e^{i\lambda}\Phi_{-}r_{L'})^{-1}(e^{i\lambda}\Phi_{-}r_{R}\Phi_{+}t_{L}a_{L}+e^{i\lambda/2}\Phi_{-}t_{R'}a_{R}).\label{eq:al-solution-HOTI}
\end{equation}
Plugging Eq.\ \eqref{eq:al-solution-HOTI} into Eq.\ \eqref{eq:1-1},
we find $b_{L}$ as
\begin{align}
b_{L} & =r_{L}a_{L}+t_{L'}(1-e^{i\lambda}\Phi_{-}r_{L'})^{-1}(e^{i\lambda}\Phi_{-}r_{R}\Phi_{+}t_{L}a_{L}+e^{i\lambda/2}\Phi_{-}t_{R'}a_{R})\nonumber \\
 & =[r_{L}+e^{i\lambda}t_{L'}(1-e^{i\lambda}\Phi_{-}r_{L'})^{-1}\Phi_{-}r_{R}\Phi_{+}t_{L}]a_{L}+e^{i\lambda/2}t_{L'}(1-e^{i\lambda}\Phi_{-}r_{L'})^{-1}\Phi_{-}t_{R'}a_{R}.\label{eq:result2}
\end{align}
Rewriting Eqs.\ \eqref{eq:result1} and \eqref{eq:result2}, we obtain
the effective scattering matrix of the junction
\begin{align}
\begin{pmatrix}b_{L}\\
b_{R}
\end{pmatrix} & =S\begin{pmatrix}a_{L}\\
a_{R}
\end{pmatrix},\ \ S=\begin{pmatrix}r & t'\\
t & r'
\end{pmatrix},\label{eq:S-matrix-1}
\end{align}
where
\begin{align}
t & =t_{R}\mathcal{S}t_{L},\ \ r'=r_{R'}+e^{i\lambda/2}t_{R}\mathcal{S}r_{L'}\Phi_{-}t_{R'},\nonumber \\
r & =r_{L}+e^{i\lambda/2}t_{L'}\mathbb{\mathcal{S}}'r_{R}\Phi_{+}t_{L},\ \ t'=t_{L'}\mathbb{\mathcal{S}}'t_{R'},\nonumber \\
\mathbb{\mathcal{S}} & =e^{i\lambda/2}\Phi_{+}(1-e^{i\lambda}r_{L'}\Phi_{-}r_{R}\Phi_{+})^{-1},\nonumber \\
\mathbb{\mathcal{S}}' & =e^{i\lambda/2}(1-e^{i\lambda}\Phi_{-}r_{L'})^{-1}\Phi_{-}.
\end{align}
 With this general scattering matrix, the two-terminal conductance
can be written as
\begin{equation}
G(B,\theta)=\frac{e^{2}}{h}\mathrm{tr}(tt^{\dagger})=\frac{e^{2}}{h}\mathrm{tr}(t_{R}^{\dagger}t_{R}\mathcal{S}t_{L}t_{L}^{\dagger}\mathcal{S}^{\dagger}).
\end{equation}

\section{Chiral hinge states under magnetic fields}

In this section, we demonstrate the splitting behavior of the chiral
hinge states when rotating the magnetic field. In the absence of magnetic
fields, the four chiral hinge states have a double degenerate linear
spectrum in $k_{z}$-direction. The left-moving hinge states cross
with the right-moving ones at $k_{z}=0$. The magnetic field gives
rise to a spatially varying vector potential. Remember that the hinge
states are localized at different hinges of the system. The local
vector potential splits the linear spectrum of hinge states. Under
the chosen gauge for the vector potential, ${\bf A}=(0,0,B[y\cos\theta-x\sin\theta])$,
the spectrum of hinge states is split as $+v(k_{z}\pm\delta k_{1}^{z})$
and $-v(k_{z}\pm\delta k_{2}^{z})$, where the splitting strengths
are determined by $\delta k_{1}^{z}=BW_{x}|\sin(\theta-\pi/4)/2|$
and $\delta k_{2}^{z}=BW_{y}|\cos(\theta+\pi/4)/2|$. Thus, the splitting
of hinge state spectrum strongly depends on the field direction $\theta$.

We focus on the splitting of the chiral hinge states in the lower-energy
spectrum presented in Fig.\ \ref{fig:splitting}. At $\theta=0\pi,$
two pairs of the hinge states split by equal value. At $\theta=\pi/4$,
only one pair of the hinge states can be split, whereas the other
one remain unaltered, i.e., $\delta k_{1}^{z1}=0$. The spectrum at
$\theta=\pi/2$ looks the same as for $\theta=0$. Another special
field direction is at $\theta=3\pi/4$ at which we have instead $\delta k_{2}^{z}=0$.
At $\theta=\pi$, the spectrum is the same as that at $\theta=0$.
This evolution with rotating the magnetic field is consistent with
the analytical results $\delta k_{1}^{z}=BW_{x}|\sin(\theta-\pi/4)/2|$
and $\delta k_{2}^{z}=BW_{y}|\cos(\theta+\pi/4)/2|$.

\begin{figure}
\centering

\includegraphics[clip,width=0.8\columnwidth]{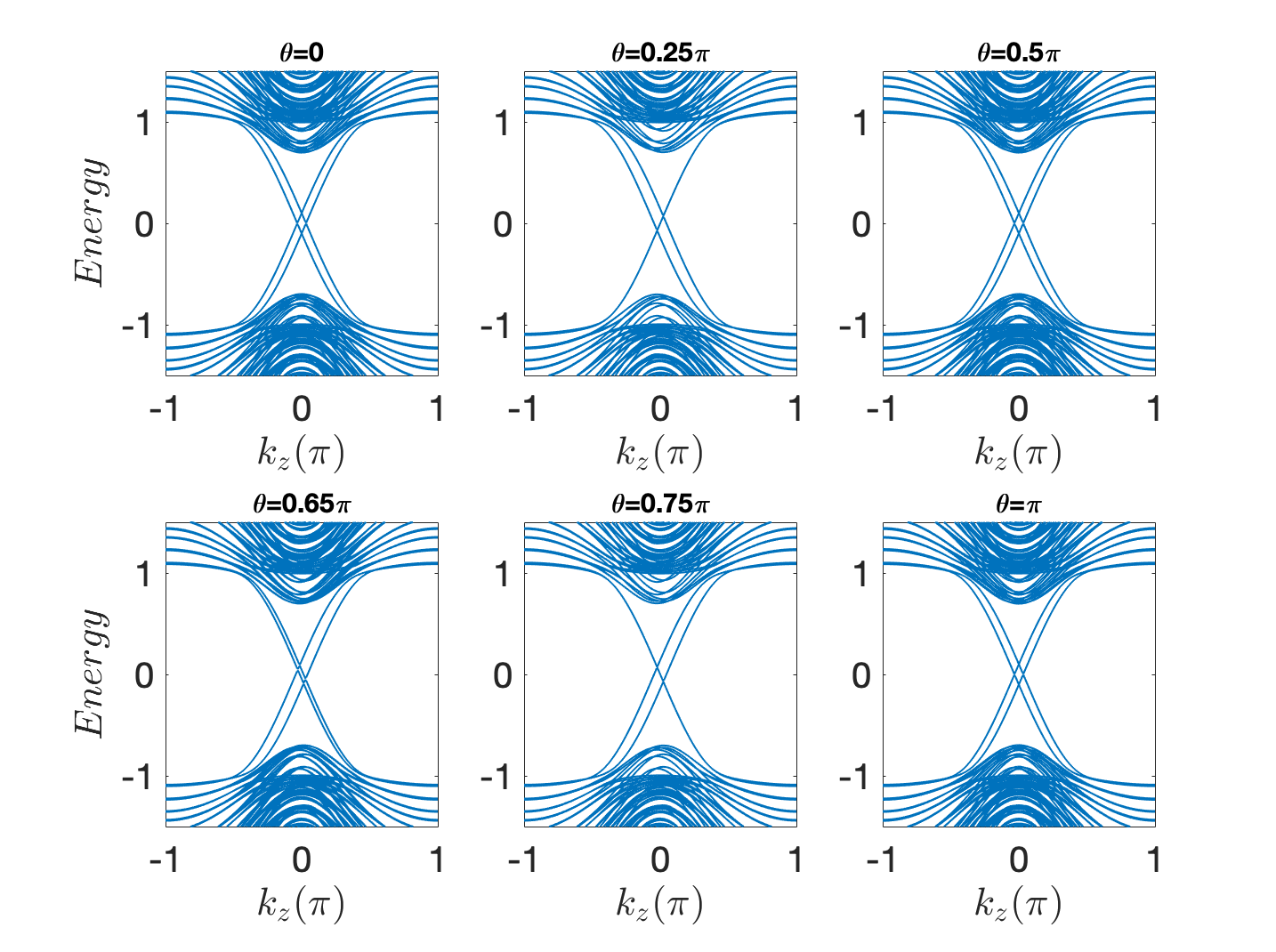}

\caption{Evolution of the hinge states spectrum when rotating the field direction
from $\theta=0$ to $\pi$. Here, we choose parameters: $L_{z}=60a$,
$W_{x}=W_{y}=12a$, $m=2,b=-1,v=1$, $\Delta=1$. The field strength
is fixed at $B=2B_{0}$.}

\label{fig:splitting}
\end{figure}

\section{Numerical simulation details}

To calculate the conductance, we employ the Landauer-B\"uttiker formalism
\citep{Landauer70pm,buttiker_four-terminal_1986,Dattabook} in combination
with lattice Green functions. The two-terminal conductance is evaluated
as
\begin{equation}
G=\frac{e^{2}}{h}\mathrm{Tr}[\Gamma_{L}G^{r}\Gamma_{R}G^{a}],
\end{equation}
where the line width function
\begin{equation}
\Gamma_{\beta}=i[\Sigma_{\beta}-\Sigma_{\beta}^{\dagger}]
\end{equation}
with the $\Sigma_{\beta}$ being the self-energy due to coupling of
the lead $\beta\in\{L,R\}$ to the central region of interest. The
retarded and advanced Green function, $G^{r}$ and $G^{a}$, are obtained
as
\begin{equation}
G^{r}=(G^{a})^{\dagger}=(E_{F}-H_{c}-\Sigma_{L}^{r}-\Sigma_{R}^{r})^{-1}.
\end{equation}
Here, both the self-energy $\Sigma_{\beta}$ and the Green function
$G^{r/a}$ can be calculated by using the recursive method \citep{MacKinnon85zpbc}.

In the numerical simulations, we choose the parameters $L_{z}=60a$,
$W_{x}=W_{y}=12a$, $m=2,$$b=-1$, $v=1$ and $\Delta=1$ for the
SOHI, and $m=3,$ $v=0,$ $b=-1$, and $\Delta=0$, and chemical potential
$\mu=-0.1$ for the two leads. Without loss of generality, we set
the Fermi energy in the SOHI at $E_{F}=0.002$.

\section{Trivial cases of perfect transmission}

In this section, we demonstrate the behaviors when the interfaces
of the proposed setup are totally transparent. Under this condition,
the chiral hinge states do not talk to each other and thus no interference
loop is forming. As a result, the two-terminal conductance $G$ is
quantized at $2e^{2}/h$ and independent of the magnetic field. Let
us consider two scenarios responsible for such transparent interfaces:
\begin{itemize}
\item Case 1: the leads are also made of the same SOTIs. Chiral hinge states
exist in all regions of space and pass from one lead to the other
lead directly;
\item Case 2: the leads are made of conventional semiconductors or topological
insulators but highly doped. In this case, there are too many channels
in the leads such that chiral hinge states loose quantum coherence
once entering the leads.
\end{itemize}
The results for these two cases are presented in Fig.\ \ref{fig:trivialcase}.
The two-terminal conductance is fixed at $2e^{2}/h$ and has no dependence
on neither the field strength $B$ nor the direction $\theta$.

\begin{figure}
\centering

\includegraphics[clip,width=0.5\columnwidth]{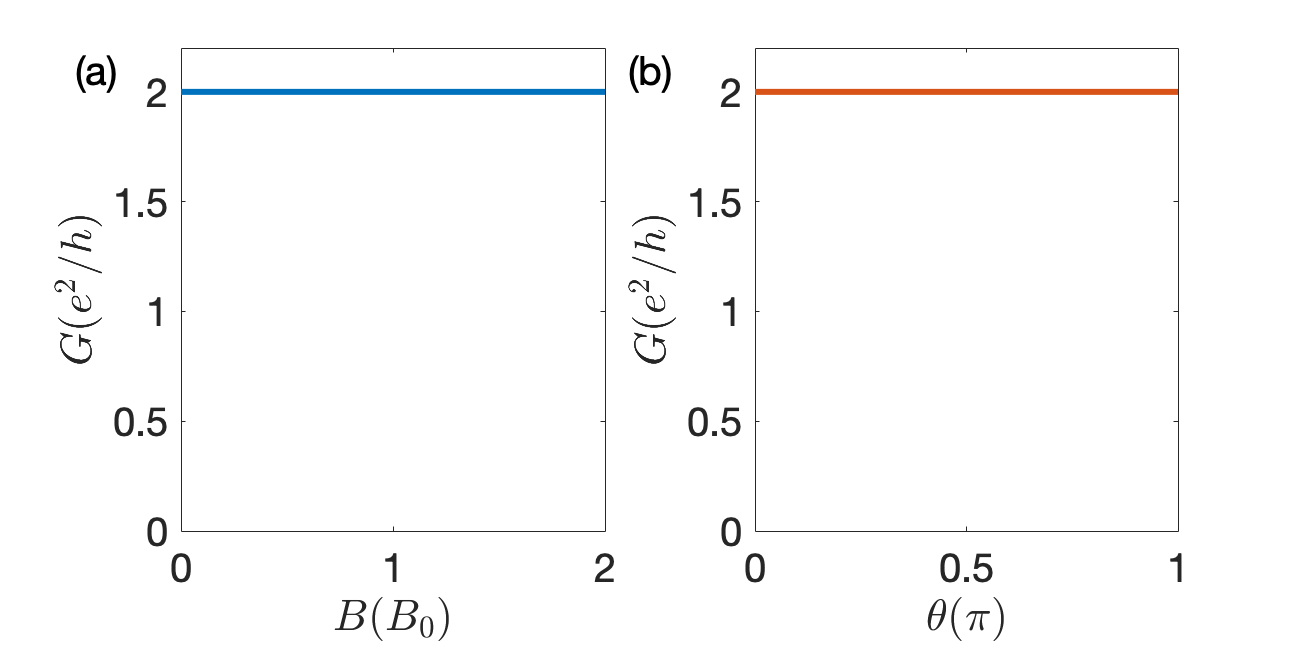}

\caption{For the two trivial cases 1 (a) and 2 (b) as discussed in this section,
the conductance is quantized at $2e^{2}/h$ and has no dependence
on either the field strength $B$ or field direction $\theta$. }

\label{fig:trivialcase}
\end{figure}

\section{Parametrizing the scattering matrix}

In this section, we parameterize the scattering matrix with the help
of the numerical method. There are two interfaces in our proposed
setup. Each interface is described by a $4\times4$ scattering matrix,
i.e., $S_{L}$ and $S_{R}$ as listed above, respectively. Due to
time-reversal symmetry breaking, $S_{L}$ and $S_{R}$ are unitary
matrices.%
Parametrizing these scattering matrices is cumbersome because of
the choice of at least 16 free parameters. Instead, we obtain the
scattering matrices directly from numerical simulations as explained
below.

\begin{figure}
\centering

\includegraphics[clip,width=0.6\columnwidth]{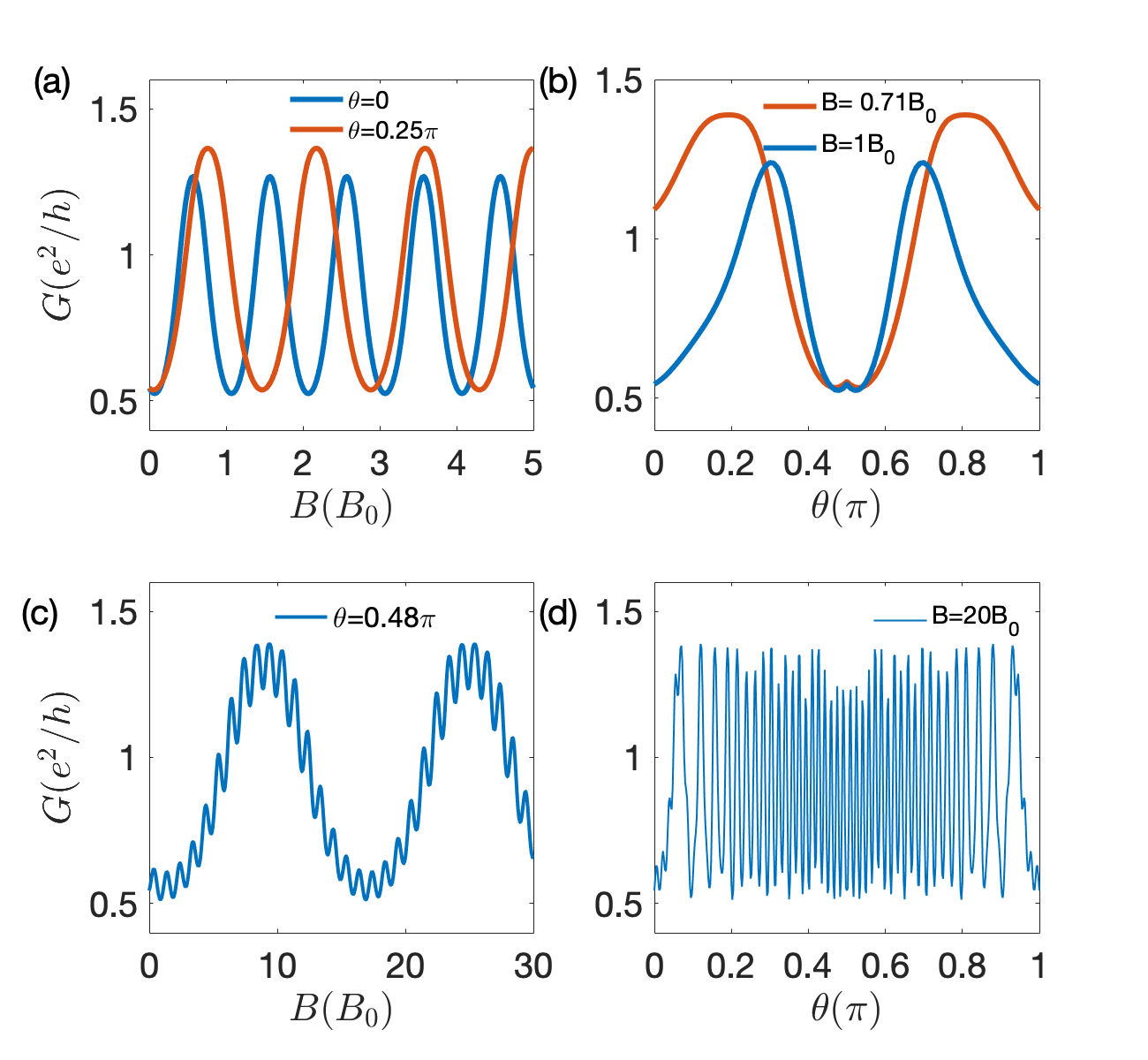}

\caption{Transport properties obtained using the analytical formula, Eq. (3)
in the main text, after we parametrize the relevant scattering matrix
according to the corresponding parameter settings. (a) Conductance
oscillation pattern as function of field strength $B$ for different
field directions $\theta=0$ and $\theta=0.25\pi$, respectively.
(b) Conductance oscillation pattern as function of field direction
$\theta$ for different field strengths $B=B_{0}/\sqrt{2}\thickapprox0.71B_{0}$
and $B=1B_{0}$, respectively. (c) Beating pattern of conductance
as function of field strength $B$. (d) Beating pattern of conductance
as function of field direction $\theta$.}

\label{fig:scatteringmatrix}
\end{figure}

Let us consider a simpler junction with two semi-infinite regions
in $z$ direction: one region is made of a trivial insulators in the
region $z<0$, and the other regions made of the SOTI in the region
$z>0$. The parameters for lead and SOTI are taken the same as those
in our interference setup. Then, the interface of this simpler setup
mimics the left interface of our interference setup. The scattering
matrix at this interface can be obtained numerically by calculating
the retarded Green functions for the two regions and then employing
the Fisher-Lee relation, or directly using the
Kwant algorithm \citep{Kwant_njp14}. A similar procedure applies
for the right interface.

Known from the conductance, described by Eqs.\ (2) and (3) in the
main text, the relevant four matrices are $t_{L},t_{R},r_{L'}$ and
$r_{R}$ (or another group $t_{L},t_{R},r_{L}$ and $r_{R'}$ ). Under
the same parameter setting with the original setup, we obtain the
four matrices as
\begin{alignat}{1}
t_{L} & =\left(\begin{array}{cc}
-0.06198685+0.18465009i, & -0.7498761-0.16515456i\\
0.60757993+0.469507i, & 0.10194116-0.16596995i
\end{array}\right),\nonumber \\
t_{R} & =\left(\begin{array}{cc}
0.55777311+0.34346096i, & -0.0156968+0.44520283i\\
-0.26134039-0.36076744i, & -0.27139131+0.59617366i
\end{array}\right),\nonumber \\
r_{L'} & =\left(\begin{array}{cc}
0.17964098-0.37617273i, & 0.39605429+0.20453843i\\
0.25086872-0.36845604i, & -0.33709527-0.2452419i
\end{array}\right),\nonumber \\
r_{R} & =\left(\begin{array}{cc}
-0.40202362-0.44148246i, & -0.06268933-0.1095996i\\
0.05582996+0.1132477i, & -0.14131563-0.58013761i
\end{array}\right).
\end{alignat}

Figure \ref{fig:scatteringmatrix} presents the transport properties
obtained using the analytical formula Eq.\ (3) in the main text after
we parameterize the relevant scattering matrix according to the corresponding
parameter settings. We see that the main features of the conductance
are qualitatively the same as the numerical results in Fig. 2 of the
main text. As shown in Fig. \ref{fig:scatteringmatrix} (a), each
pattern has single frequency; the oscillation amplitude is larger
at $\theta=\pi/4$; and the period of red line is about $\sqrt{2}$
times that of the blue line. In Fig. \ref{fig:scatteringmatrix} (b),
there are two peaks and the oscillation amplitude is more pronounced
when $B=B_{0}/\sqrt{2}\thickapprox0.71B_{0}$. In Fig. \ref{fig:scatteringmatrix}
(c), the conductance shows a beating pattern of $B$ at $\theta=0.48\pi$.
Finally, in Fig. \ref{fig:scatteringmatrix} (c), the conductance
shows an ``irregular'' beating pattern as a function of $\theta$
for large $B$.

\section{Model of $\mathcal{C}_{6}$ symmetric SOTIs}

The effective model for a $\mathcal{C}_{6}$ symmetric SOTI on a stacked
hexagonal lattice can be written as \citep{ZhangRX20prl}
\begin{equation}
H_{\text{hex}}=\begin{pmatrix}h+ms_{z}\sigma_{0} & h_{AB}\\
h_{AB}^{\dagger} & h+ms_{z}\sigma_{0}
\end{pmatrix},
\end{equation}
where
\begin{align}
h & =\tilde{C}-\dfrac{4}{3}C_{2}(\cos k_{1}+\cos k_{2}+\cos k_{3})+\dfrac{v}{3}(2\sin k_{1}+\sin k_{2}+\sin k_{3})\Gamma_{1}\nonumber \\
 & \ \ \ +\dfrac{v}{\sqrt{3}}(\sin k_{2}-\sin k_{3})\Gamma_{2}+w[-\sin k_{1}+\sin k_{2}+\sin k_{3}]\Gamma_{4}+\Big[M-\dfrac{4}{3}M_{2}(\cos k_{1}+\cos k_{2}+\cos k_{3})\Big]\Gamma_{5}\nonumber \\
h_{AB} & =-2C_{1}\cos k_{z}+2v_{z}\sin k_{z}\Gamma_{3}-2M_{1}\cos k_{z}\Gamma_{5},
\end{align}
and $k_{1}=k_{x},$ $k_{2}=(k_{x}+\sqrt{3}k_{y})/2$ and $k_{3}=k_{1}-k_{2}$.
 $h_{AB}$ describes the hopping between neighboring layers. The $\Gamma$
matrices are defined as $\Gamma_{i}=s_{i}\sigma_{1}$ with $i\in\{1,2,3\}$,
$\Gamma_{4}=s_{0}\sigma_{2}$ and $\Gamma_{5}=s_{0}\sigma_{3}$ with
${\bf s}$ and ${\bf \sigma}$ the Pauli matrices and $s_{0}$ and
$\sigma_{0}$ the corresponding identity matrices. The model parameters
are defined as $\tilde{C}=C_{0}+2C_{1}+4C_{2}$, and $\tilde{M}=M_{0}+2M_{1}+4M_{2}$.
In the numerical calculations, we set the parameters $C_{0}=0,$ $C_{1}=C_{2}=0.5,$
$M_{0}=-2.5$, $M_{1}=M_{2}=1,$ $v=v_{z}=1$, $m=0.5$ and $w=2$.
The perimeter of the hexagonal prisms is $6\times n_{s}$ with side
length $n_{s}=25a$ and $L_{z}=50a$.

\begin{figure}
\centering

\includegraphics[clip,width=0.6\columnwidth]{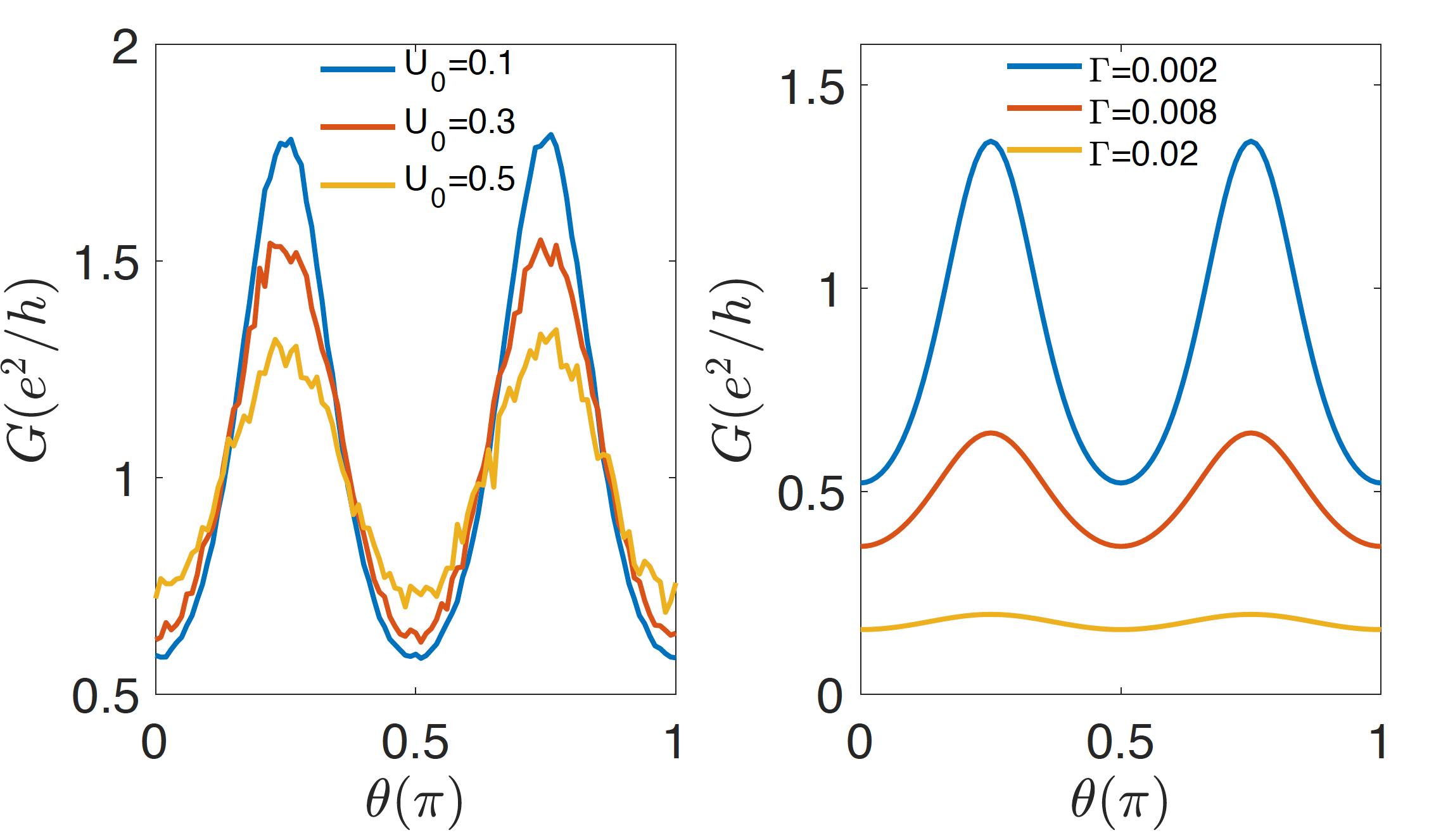}

\caption{Left panel: influence of disorder on the interference pattern. We
average over 100 disorder configurations. Right panel: influence of
dephasing on the interference pattern. We choose parameters: $L_{z}=60a$,
$W_{x}=W_{y}=12a$, $m=2,b=-1,v=1$, $\Delta=1$. The magnetic field
strength is fixed at $B=B_{0}/\sqrt{2}\thickapprox0.71B_{0}$.}

\label{fig:disorder}
\end{figure}

\begin{figure}
\centering

\includegraphics[clip,width=0.9\columnwidth]{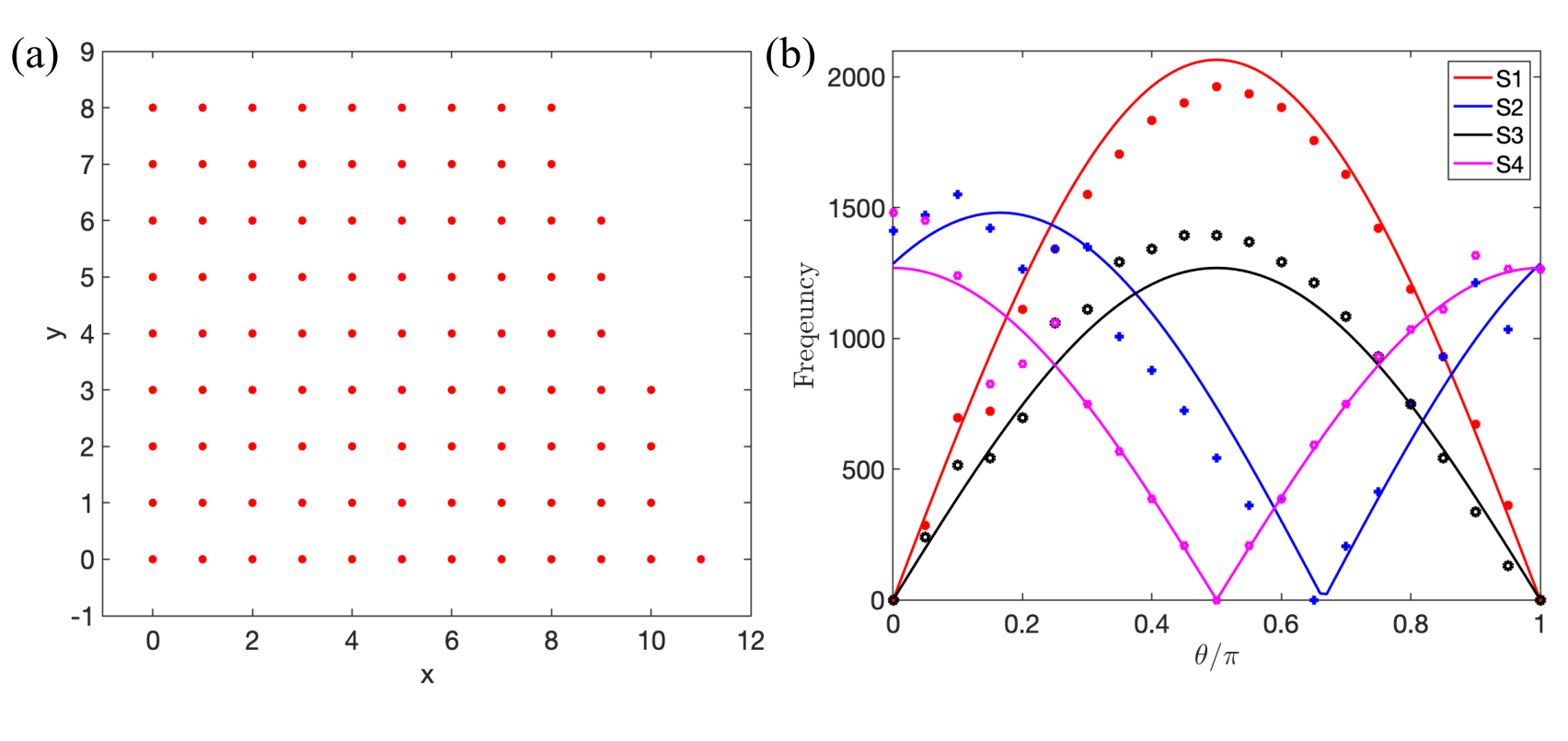}

\caption{(a) Cross section of the SOTI in a trapezoid geometry. (b) Four frequencies
as functions of field direction $\theta$. Here, S1 (S2, S3, S4) indicates
the bottom (right-side, top, left-side) surface of the trapezoid.
Solid lines are analytical results, and the dotted lines are obtained
by Fourier transformation from the conductance beating pattern. Discrepancy
between them maybe due to  the finite-size effects. We choose parameters
for the SOTI as: $L_{z}=200a$, $m=2,b=-1,v=1$, and $\Delta=1$.}

\label{fig:trapezoid}
\end{figure}

\section{Disorder and dephasing }

In this section, we show that the interference pattern of our interferometer
is robust against disorder and dephasing.

To mimic disorder, we consider the onsite type $V_{\mathrm{dis}}=V({\bf r})I_{4\times4}$
with random function $V({\bf r})$ distributed uniformly within the
interval $[-U_{0}/2,U_{0}/2]$ and $U_{0}$ being the disorder strength.
It is shown in Fig.\ \ref{fig:disorder} that as increasing the disorder
strength $U_{0}$, the oscillation amplitude decreases gradually.
However, the interference pattern of the conductance remains even
when the disorder strength is quite strong (comparable with the bulk
gap).

We also consider dephasing in the SOTI region in our setup by attaching
each site in the discretized lattice model with a virtual lead \citep{JiangH09prl}.
These virtual leads are coupled to the system via the self-energy
$-i\Gamma/2$ with $\Gamma$ measuring the dephasing strength ($1/\Gamma$
signifies the quasiparticle life time). It is shown in Fig.\ \ref{fig:disorder}
that the interference pattern of the conductance remains under weak
dephasing strength. As increasing dephasing strength $\Gamma$, the
electrons loose their phase memory quickly and thus the oscillation
amplitudes decrease accordingly.

The above results shows the oscillation pattern basically remains
under weak disorder and dephasing, which indicates the robustness
of our proposal to show quantum interference of hinge states.

\section{Multiple frequencies when the cross section is a trapezoid}

In this section, we present the multiple-frequency case when the cross
section of SOTI is a trapezoid, as shown in Fig.\ \ref{fig:trapezoid}(a).
In this case, there are generally four frequencies in the conductance
oscillation as function of field strength $B$. Figure\ \ref{fig:trapezoid}(b)
shows the four frequencies as varying the field direction $\theta$.
The numerical results are basically consistent with the analytic ones
(obtained by the effective surface areas).

One can see from Fig.\ \ref{fig:trapezoid}(b) that the four frequencies
do not increase or decrease simultaneously, which stems from the 3D
nature of the interferometer, as discussed in the main text. Beside,
we find that there always exist two field directions (the field directions
along the diagonal of the trapezoid), at which only two of the four
frequencies will survive.

\end{widetext}

\end{document}